\documentclass[aps,prl,twocolumn,superscriptaddress]{revtex4-1}
\usepackage{booktabs}
\usepackage{mathrsfs,amssymb,graphics,subfigure,threeparttable}
\usepackage{graphicx}
\usepackage{epstopdf}
\usepackage{amssymb}
\usepackage{enumerate}
\usepackage{amsmath}
\usepackage{fancyhdr}
\usepackage{bm}
\usepackage{xcolor}

\usepackage[squaren]{SIunits}
\usepackage{url}
\PassOptionsToPackage{hyphens,spaces,obeyspaces}{url}
\usepackage[hidelinks,breaklinks]{hyperref}

\begin{document}

%\preprint{}

\title{A fully plasma based electron injector for a linear collider or XFEL
}

\author{T. N. Dalichaouch}
\thanks{tdalichaouch@gmail.com}
\affiliation{Department of Physics and Astronomy, University of California, Los Angeles, California 90095, USA}

\author{ X. L. Xu}
\affiliation{State Key Laboratory of Nuclear Physics and Technology, and Key Laboratory of HEDP of the Ministry of Education, CAPT, Peking University, Beijing, 100871, China}
\affiliation{Beijing Laser Acceleration Innovation Center, Beijing 100871, China}

\author{F. Li}
\affiliation{Department of Engineering Physics, Tsinghua University, Beijing 100084, China}
\author{F. S. Tsung}
\affiliation{Department of Physics and Astronomy, University of California, Los Angeles, California 90095, USA}
\author{W. B. Mori}
\affiliation{Department of Physics and Astronomy, University of California, Los Angeles, California 90095, USA}
\affiliation{Department of Engineering, University of California, Los Angeles, California 90095, USA}

\date{\today}

\begin{abstract}
We demonstrate through high-fidelity particle-in-cell simulations a simple approach for efficiently generating 20+ GeV electron beams with the necessary charge, energy spread, and emittance for use as the injector for an electron arm of a future linear collider or a next generation XFEL.  The self-focusing of an unmatched, relatively low quality, drive beam results in self-injection by elongating the wakefield excited in the nonlinear blowout regime. Over pump depletion distances, the drive beam dynamics and self-loading from the injected beam leads to extremely high quality and high energy output beams. For plasma densities of  $10^{18}  ~\centi\meter^{-3}$, PIC simulation results indicate that self-injected beams with $0.52 ~\nano \coulomb$ of charge can be accelerated to $\sim 20$ GeV energies with projected energy spreads, $\lesssim 1\%$ within the beam core, slice normalized emittances as low as $110 ~\nano\meter$, a peak normalized brightness $\gtrsim 10^{19}~\ampere/\meter^2/\rad^2$, and  energy transfer efficiencies $\gtrsim 54\%$.    \end{abstract}

\pacs{}
%\keywords{}

%TODO
\maketitle
%%%%%%%%

Plasma-based acceleration (PBA) \cite{tajimadawson,chendawson1985} is a promising avenue for high gradient ($> 100$ GV/m) acceleration \cite{hoganplasma4Gev10cm, blumenthal42GeV85cm, leemansplasma4GeV9cm, leemans20061GeV3cm, wang20132GeV7cm, hafz2008GeVbunches,litos2014high,adli2018acceleration,steinke2016multistage,PhysRevLett.122.084801} and high quality beam generation \cite{ionizationinjconcept2006, ionizationinjconcept2006-2, ionizationinjconcept2006-3,xu2014colinearionization, li2013transversecollid, twocolorionization2014,PhysRevSTAB.17.061301,2012hiddingbeamlaserionization, bulanovdownramp1998,sukdownramp2001, fubiani06,xu2017downrampinj,martinezdownramp2017,xu2020generation,geddes08, gonsalvesdownramp2011,buck13, kalmykov09, xu2005tightfocusedlaser, Thamine2020, xu2023lwfageneration,li2022ultrabright}. PBA could thus be the mechanism to provide an injector for next generation x-ray light sources \cite{xfelbarletta2010,27nmFELnature2021} 
or a future linear collider (LC). For an x-ray free-electron laser (XFEL), the generation of high power ($> 100$ GW) requires GeV-class beams with high currents $I \gtrsim 10$ kA, low normalized emittances $\epsilon \lesssim 100$ nm, and low energy spreads $\sigma_{\gamma} \lesssim 1\%$ \cite{xfelbarletta2010}. These beams are typically characterized by high peak normalized brightnesses $B_n  = 2I/\epsilon_n^2 \gtrsim 10^{18}\ \ampere/\meter^2/\radian^2$. Similarly, designs for TeV-class linear colliders also require low emittance, high charge particle beams to achieve high luminosities for e-e+ beam collisions at the interaction point (IP). 

To produce the desired beams needed for LC and XFEL applications using PBA, numerous injection schemes \cite{ionizationinjconcept2006, ionizationinjconcept2006-2, ionizationinjconcept2006-3,xu2014colinearionization, li2013transversecollid, twocolorionization2014, PhysRevSTAB.17.061301, 2012hiddingbeamlaserionization,bulanovdownramp1998,sukdownramp2001, fubiani06,xu2017downrampinj,martinezdownramp2017,geddes08, gonsalvesdownramp2011,buck13, xu2020generation,kalmykov09, xu2005tightfocusedlaser, Thamine2020, xu2023lwfageneration,li2022ultrabright} have been proposed in which plasma electrons are trapped and accelerated by nonlinear wakefields in the blowout regime. In these nonlinear wakes, plasma electrons are fully blown out by the intense fields of a laser pulse or particle beam driver creating an ion channel surrounded by a thin sheath of electrons. These sheath electrons can typically propagate at velocities near the speed of light at the rear of the channel and are therefore natural candidates to be trapped if the phase velocity,
$v_{\phi}$, can be controlled using a plasma density down ramp (DDR) \cite{bulanovdownramp1998,sukdownramp2001, fubiani06,xu2017downrampinj,martinezdownramp2017,geddes08, gonsalvesdownramp2011,buck13, xu2020generation}, an evolving driver \cite{kalmykov09, xu2005tightfocusedlaser, Thamine2020, xu2023lwfageneration}, or flying focus \cite{li2022ultrabright}. Particle trapping induced by wakefield elongation has been shown to produce narrow beams with normalized emittances several orders of magnitude smaller than those of the drive beams \cite{xu2017downrampinj}. 

Maximizing energy gain and minimizing energy spread is challenging because it requires optimal beam loading \cite{tzoufrasprl, tzoufrasprab, Thamine2021,shiyu22} over pump depletion distances. While beams with nearly trapezoidal current profiles can be used to load constant wakefields \cite{tzoufrasprl, tzoufrasprab, Thamine2021, shiyu22}, they can be difficult to realize through self-injection. Recently, low energy spread PBA schemes \cite{BO2, liu2023} facilitated by a dynamic beam loading (DBL) effect in Bayesian-optimized or tri-plateau plasma density profiles \cite{chiou1998,liu2023} have attracted significant interest because they do not require such shaped bunches.

\begin{figure}[t]
\centering
\includegraphics[width=0.5\textwidth]{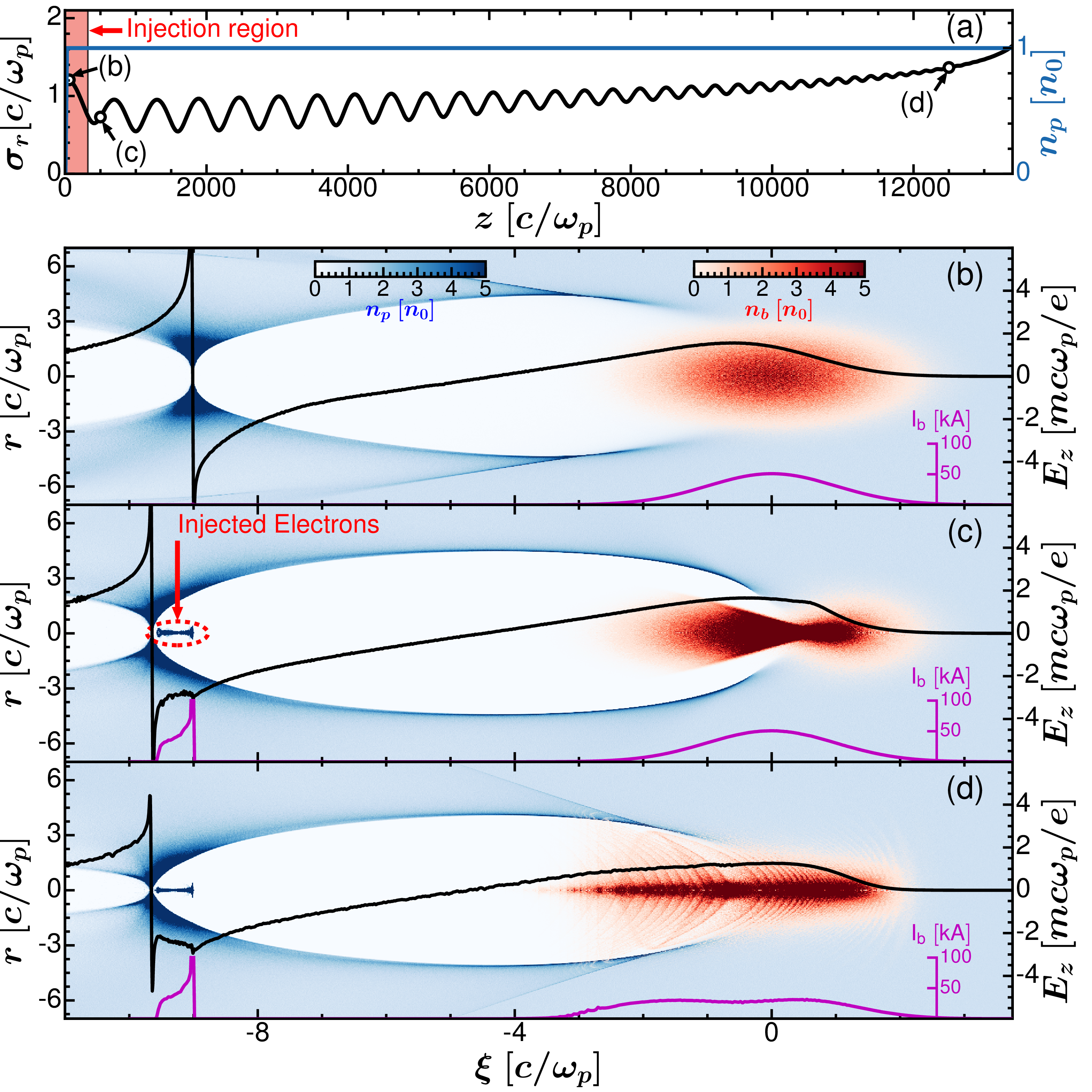}
\caption{\label{fig:rotation_schematic} (a) Simulated spot size evolution $\sigma_r(z)$ (black) of an electron driver $\{ \gamma_b = 20000, \Lambda = 6, k_p\sigma_z = 1, k_p\sigma_0 = 0.5\sqrt{\Lambda} , k_{\beta}\beta^* = 16 \}$ propagating in a constant plasma density (blue). Snapshots of the electron charge density distribution and axial electric field $E_z(\xi)$ (black) are shown at (b) $k_pz= 60$, (c) $k_pz = 500$, and (d) $k_pz=12500 $. Current profiles of the drive and injected beam are plotted (purple). }
\end{figure}

In this Letter, we present a 
novel approach to high energy, high brightness, low energy spread beam generation driven by DBL \cite{chiou1998} in a self-injected, high efficiency, constant density PBA stage. 
An unmatched electron driver initially self-focuses its own wakefield leading to self-injection \cite{Thamine2020}. A DBL effect is subsequently induced by the plasma wake evolution over the pump depletion length. Using particle-in-cell (PIC) simulations and theory, we show that the wake evolution is caused by spot size expansion and dephasing of the driver due to pump depletion and betatron oscillations in the plasma. The electric field chirp $d_{\xi}E_z$ within the trailing beam core decreases from positive (underloaded) to negative (overloaded) due to pump depletion, where $\xi \equiv z-ct$ is the co-moving coordinate. The cumulative effect is a trailing beam with small energy spread near pump depletion.

The physics of the DBL described here is different from those described in Refs.~\citenum{BO2,liu2023,chiou1998} that rely on dephasing in hollow channels or phase advancement from a discrete or continuous set of density plateaus within laser driven wakefields.  For beam-driven wakes considered here, the wake transitions from underloaded to overloaded due to a decrease in the wake strength as the driver shape evolves due to pump depletion.

Based on PIC simulations using {\scshape osiris} \cite{osiris}, we find that for parameters that may be possible at FACET II \cite{facet2016} high quality trailing beams can be injected and efficiently accelerated to energies up to $20$ GeV with low normalized core energy spreads $\hat{\sigma}_{\gamma} \lesssim 1\%$ and high normalized brightnesses $B_n \gtrsim 10^{19} \ampere/\meter^2/\radian^2$. The process is robust and works for Gaussian drivers with different durations $\sigma_z$.  Simulation results presented in this Letter use the quasi-3D algorithm implemented in {\scshape osiris} \cite{osiris} with high resolution grids $\Delta z = \Delta r = \frac{1}{128} \frac{c}{\omega_p}$, $\Delta t = \frac{1}{512}\frac{1}{\omega_p}$, where $\omega_p ^2= \frac{4\pi e^2n_p}{m_e}$ is the plasma frequency. The algorithm expands the fields into azimuthal modes (m). In the full length simulations only the $m=0$ mode is kept while simulations of the initial injection and acceleration phase with $m$ up to 2 were carried out. A customized finite-difference solver and current deposit is used to reduce numerical effects due to injection \cite{xu2013numerical, xu2020,li2017}.

The key physics of this process is illustrated in Fig.~\ref{fig:rotation_schematic}. A 10 GeV ($\gamma_b = 20000$) electron driver with a peak current of 51 kA ($\Lambda  \equiv \frac{\omega_p^2}{n_p c^2} \int_{0}^{r \gg \sigma_r} n_b rdr = 6$) is initialized at the entrance of a constant density plasma (blue) with a focal spot size $k_p\sigma_0 \approx 1.225$, $k_p\sigma_z = 1$, and Courant-Snyder (C-S) \cite{csparams} parameters $\beta_i = \langle x^2 \rangle /\epsilon \approx 3200 \ c/\omega_p$ and $\alpha_i = -\langle x x^{\prime} \rangle/\epsilon \approx 0$, where $k_p = \frac{\omega_p}{c}$ is the plasma wavenumber, $\epsilon = \sqrt{ \langle x^2 \rangle  \langle {x^{\prime }}^2 \rangle - {\langle xx^{\prime} \rangle}^2}$ is the geometric emittance, and $\epsilon_n \approx \gamma_b \epsilon$ is the normalized emittance. For a plasma density of $n_0 \simeq 10^{18} \ \centi\meter^{-3}$, these parameters correspond to an electron beam with charge $ Q_d \simeq 2.27 \ \nano \coulomb$, diffraction length $\beta^* \simeq 1.7 \ \centi\meter $,  spot size $\sigma_r \simeq 6.5 \ \micro \meter$, and beam length $ \sigma_z \simeq 5.31 \ \micro \meter $.

As the spot size of an electron drive beam decreases the ion channel elongates leading to self-injection \cite{Thamine2020}. This can be seen in Figs.~\ref{fig:rotation_schematic}(a)-(c). The spot size evolution of the driver 
%in the plasma 
depends on its betatron wavenumber $k_{\beta} \equiv \frac{k_p}{\sqrt{2\gamma_b}}$ and diffraction length $\beta^* \equiv \sigma_0^2/\epsilon$, where $\sigma_0$ is the focal spot size. Since the driver is not matched and $k_{\beta}\beta^* \approx 16 \gg 1$, it is self-focused by the plasma ion channel, and the projected spot size, $\sigma_r(z)$, oscillates over the length scale of the betatron wavelength $\lambda_{\beta} = 2\pi/k_{\beta}$ as shown in Fig.~\ref{fig:rotation_schematic}(a). Wake expansion therefore occurs due to spot size focusing \cite{Thamine2020} over $\lambda_{\beta}/4 \sim 314 \ c/\omega_p$, thereby reducing the phase velocity of the wake. As a result, energetic sheath electrons can be trapped and accelerated at the rear the ion channel as shown in Figs.~\ref{fig:rotation_schematic}(b)-(c). 

While $\sigma_r(z)$ continues to oscillate after the first betatron oscillation, the amplitude of the subsequent oscillations is not as large. There is phase mixing between slices within the ion channel (rear) and slices at the head where  blowout has not been fully established (smaller $k_{\beta}$). Thus injection is limited to the shaded region in Fig.~\ref{fig:rotation_schematic}(a).

The injected beam has a nearly trapezoidal current profile (purple) with high slice currents ($\sim 30-60$ kA) in Figs.~\ref{fig:rotation_schematic}(b)-(d) within the beam core; however, it does not perfectly flatten the wakefield at $z= 500 \ c/\omega_p$ [Fig.~\ref{fig:rotation_schematic}(b)]. As a result, a small positive electric field chirp $(d_{\xi}E_z > 0)$ is observed following injection. However, this chirp dynamically increases from positive (underloaded) to negative (overloaded) after the beam has propagated $z = 12500 \ c/\omega_p$ as seen in Figs.~\ref{fig:rotation_schematic}(c)-(d). As we will show, this DBL process is dictated by the drive beam energy. The wake therefore remains weakly underloaded over most of the acceleration and only becomes overloaded when the driver has significantly pump depleted ($\gtrsim 80\%$ energy loss). The injected beam is rapidly dechirped by the strong overloading (large negative $d_{\xi} E_z$) in the final stage of acceleration [Fig.~\ref{fig:rotation_schematic}(d)]. As a result, the injected beam can be extracted with a low energy spread when the driver has nearly fully pump depleted at $z = 13240 \ c/\omega_p$.

The DBL process illustrated in Fig.~\ref{fig:rotation_schematic} is effectuated by two drive beam dynamics: (i) spot size defocusing and (ii) longitudinal dephasing. The driver spot size increases and current profile elongates as it loses energy and undergoes betatron oscillations [Fig.~\ref{fig:rotation_schematic}]. This alters the wake excitation and shape of the ion channel, $r_b(\xi)$, and, thus, the loading of the channel by the injected bunch.

\begin{figure}[t]
\centering
\includegraphics[width=0.5\textwidth]{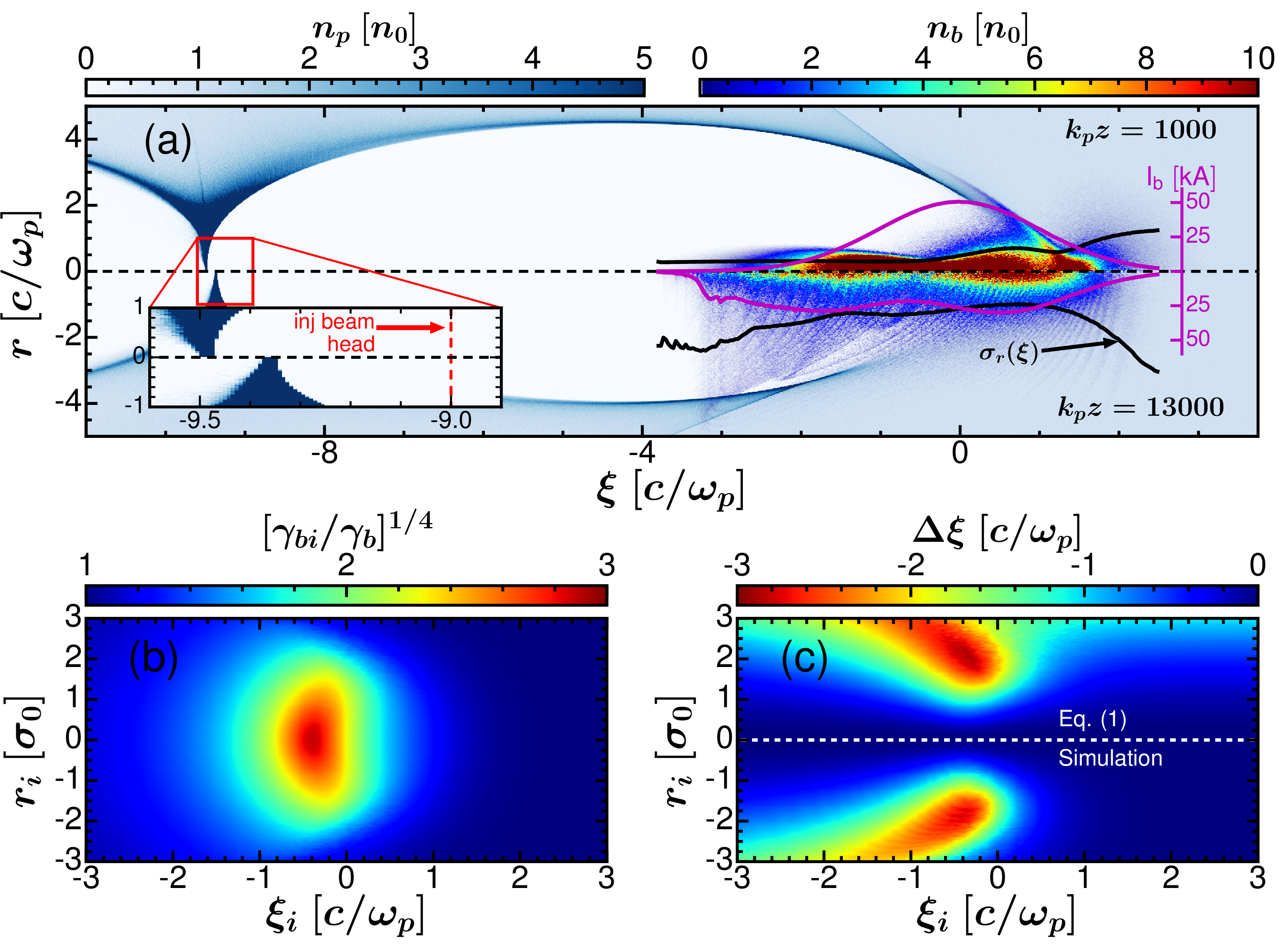}
\caption{\label{fig:beam_evolution} (a) Unloaded plasma wakes excited by the driver in Fig.~\ref{fig:rotation_schematic} at $k_pz = 1000$ (top) and $k_pz = 13000$ (bottom). Slice spot sizes $\sigma_r(\xi)$ (solid black) and current profiles (purple) are annotated. (b) $[\gamma_{bi}/\gamma_b]^{1/4}$ and (c) $\Delta\xi$ as a function of the initial beam particle position $(\xi_i,r_i)$ at $k_p z =13000$. }
\end{figure}

We performed additional PIC simulations to obtain the unloaded wakefields by using the current density profile of the drive beam taken at $k_p z = 1000$ and $k_p z = 13000$. A comparison of the results is shown in Fig.~\ref{fig:beam_evolution}(a). The bubble trajectory, $r_b(\xi)$, crosses the axis slightly sooner at $k_pz = 13000$ (see inset) due to spot size defocusing \cite{Thamine2020}. A slightly smaller blowout radius $k_p r_m \approx 2 \sqrt{\Lambda}$ \cite{lu2006nonlinearphysplasma,lu2006nonlineartheoryprl} is also produced due to the reduction in the driver peak current (purple); the double-peaked current profile arises because of slice dephasing around the driver centroid. These effects lead to a smaller $r_b$ at head of the injected beam $(k_p\xi \approx -9)$ and stronger loading from the space-charge force of the injected beam.

The transverse and longitudinal drive beam dynamics in Fig.~\ref{fig:beam_evolution}(a) can be well-understood by examining the motion of the underlying beam electrons as they execute betatron oscillations in a uniform plasma and pump deplete. For a relativistic beam electron ($\vec{v} \simeq c\hat{z}$) propagating inside the ion channel, the conservation of $\oint dx p_x$ implies the transverse coordinates follow $\boldsymbol{x}(z) = \boldsymbol{A_{i}} [\gamma_{bi}/{\gamma_b(z)}]^{1/4} \cos(\phi + \boldsymbol{\phi_i})$  \cite{xu2014colinearionization,yujianprab, ariniello2022chromatic}, 
where $\phi(z) = \int_0^z k_{\beta}(s)ds$ is the betatron phase, $\boldsymbol{A_{i}} \approx \sqrt{\boldsymbol{x_{i}}^2 + \boldsymbol{x_{i}}^{\prime 2}/k_{\beta i}^2}$, $\cos\boldsymbol{\phi_i} = \boldsymbol{x_{i}}/\boldsymbol{A_{i}}$, and subscript ``i" denotes initial values.

Defocusing is caused by the energy dependence of the betratron oscillations $\boldsymbol{x} \sim [\gamma_{bi}/\gamma_b(z)]^{1/4}$.  As the beam electrons lose energy to the wake, their betatron oscillations increase in amplitude, leading to a larger projected spot size. In Fig.~\ref{fig:beam_evolution}(b), it can be seen that the energy factor $[\gamma_{bi}/\gamma_b(z)]^{1/4}$ ranges from 2 to 3 for particles that have significantly pump depleted ($93\%+$ energy loss) at $k_pz = 13000$. These particles originate from positions where the decelerating field $E_z$ is largest (behind the beam centroid $\xi_i \approx 0$ and inside the channel $r_i < r_b$). This pump depletion effect is the predominant reason for  the increase in the slice spot size along the middle and rear of the beam in Fig.~\ref{fig:beam_evolution}(a). While spot size expansion still occurs at the beam head due to diffraction, the effect is limited to a relatively small amount of charge ($\sim 6\%$).

Longitudinal dephasing and slice mixing occurs as the beam electrons execute betatron oscillations. This is because the beam electrons do not travel in straight lines.
Electrons oscillating with a transverse velocity, $v_{\perp}$, slip backwards with $\frac{d\xi}{cdt} =\frac{v_z}{c}-1 \approx -\frac{v_{\perp}^2}{2v_z^2}$. The total dephasing over a propagation distance $z$ can be well-approximated by (see Supplemental Material),
\begin{align}
\Delta\xi(z) &=  \frac{-k_p^2|\boldsymbol{A_i}|^2 z}{4\gamma_b \left(1 + \sqrt{ \gamma_{bi}/\gamma_b}\right)}.\label{eq:dephasefinal}
\end{align}

Eq.~(\ref{eq:dephasefinal}) is evaluated at $k_p z = 13000$ in Fig.~\ref{fig:beam_evolution}(c) and exhibits good agreement with $\Delta \xi$ obtained directly from the simulation results. Upon inspection of Eq.~(\ref{eq:dephasefinal}), it is evident that particles with low energies $\gamma_b \ll \gamma_{bi}$ and large amplitudes $|\boldsymbol{A_i}| \sim r_i$ dephase the most. Such particles are initially located behind the centroid $-1 \lesssim k_p\xi_i \lesssim 0$ near $r_i \sim 2 \sigma_0$ and dephase by as much as $\Delta\xi \simeq - 2.8 \ c/\omega_p$. This dephasing effect on these particles produces the double-peaked current profile in Fig.~\ref{fig:beam_evolution}(a).

\begin{figure}[b]
 \centering
 \includegraphics[width=0.5\textwidth]{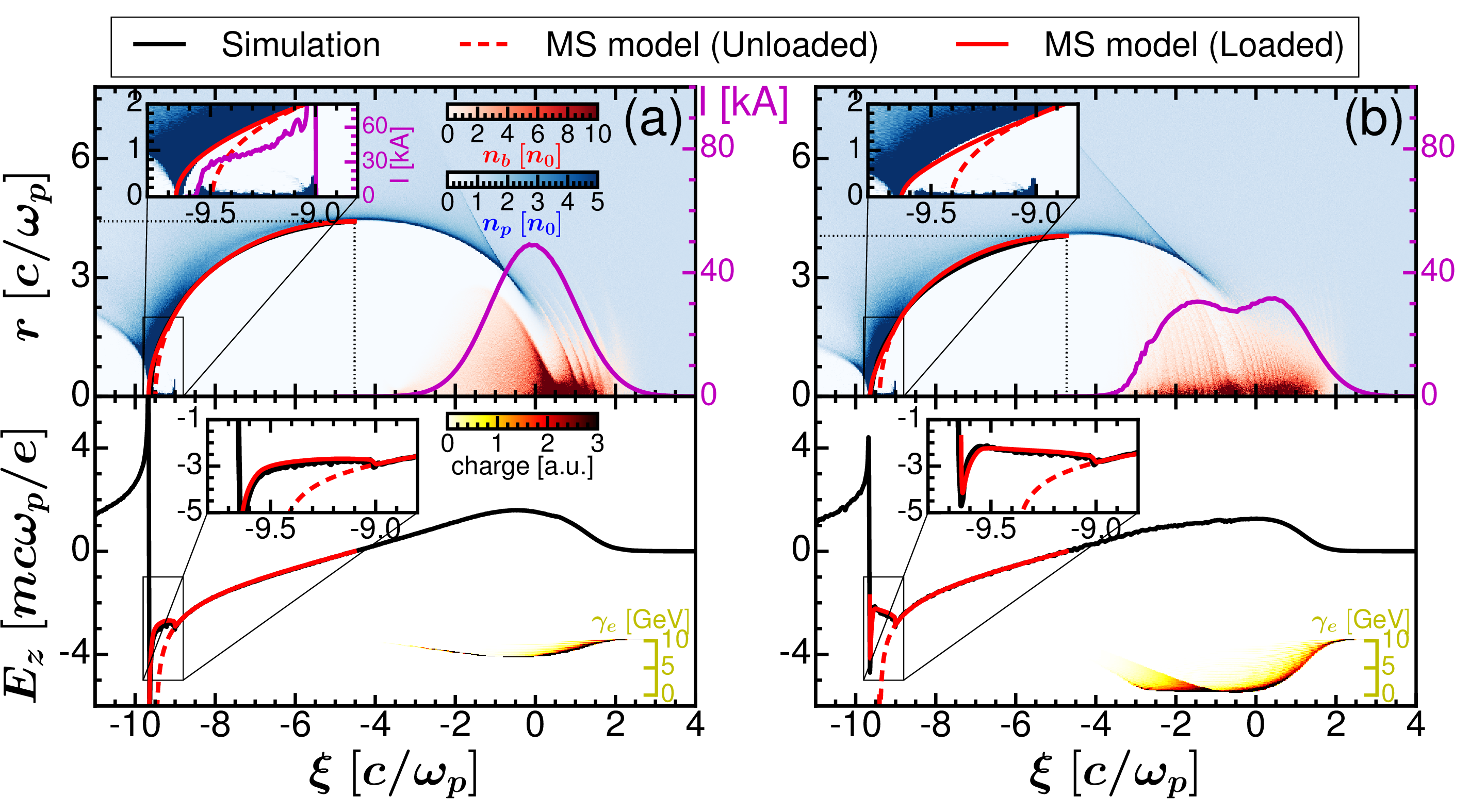}
 \caption{\label{fig:injection_optimal} The electron density distribution of the plasma and driver, axial electric field, and drive beam energy distribution at (a) $ k_pz = 4000$, and (b) $ k_pz = 12500$. Numerical calculations of $r_b(\xi)$ and $E_z(\xi)$ using the multi-sheath (MS) model are shown with (solid red) and without (dashed red) the injected bunch. Integration parameters are provided in the Supplemental Material.  
 }
 \end{figure}

 The drive beam evolution described above is inherently dictated by the beam energy $\gamma_b$. As the driver pump depletes, the change in the current density profile leads to a slightly smaller blowout radius and shorter wake wavelength when there is no beam load [Fig.~\ref{fig:beam_evolution}(a)]. The evolution of the channel shape alters how it is loaded. This DBL process is illustrated in Fig.~\ref{fig:injection_optimal} using snapshots of the plasma wake, electric field, and drive beam energy distribution at different stages of acceleration. The wakefield at each stage is fully described by the wake potential $\psi = e(\phi - A_z)/mc^2$ where $\phi$ and $A_z$ are the scalar and axial vector potential, respectively. Beam loading effects are analyzed using the multi-sheath model \cite{Thamine2021}. Inside the channel, the axial potential $\psi_0(\xi) = (1+\beta^{\prime})k_p^2r_b^2/4$ , which is used to calculate $r_b(\xi)$ and $E_z(\xi) = d_{\xi} \psi_0$. Details on the model parameters and equations are provided in the Supplemental Material.

At $z = 4000 \ c/\omega_p$, the wakefield remains slightly underloaded with a positive chirp ($d_{\xi}E_z > 0$) along the injected bunch [Fig.~\ref{fig:injection_optimal}(a)]. The ion channel shape has not changed significantly since the termination of injection. However, after significant pump depletion (half the beam has lost most of its energy), the wakefield becomes overloaded ($d_{\xi}E_z < 0$) at $z = 12500 \ c/\omega_p$ [Fig.~\ref{fig:injection_optimal}(b)]. The shift to overloaded is attributed to the evolution of the ion channel: $r_b$ is now smaller at the head of the injected beam. The innermost sheath electron thus feels a stronger space-charge force, $F_d = \frac{\lambda(\xi)}{r_b}$, as it passes the injected beam thereby reducing $|p_{\perp}|$. As the stronger loading modifies the electron momenta, it also alters the wake potential which obeys the relationship $\psi = \gamma - p_z/mc - 1$ \cite{mora,Thamine2021}. Based on PIC simulation results, we find that the minimum wake potential $\psi_{min}$ increases from $-1$ to $-0.93$ from $k_p z = 4000$ to $k_p z = 12500$. Due to these effects, the slope $\frac{dr_b}{d\xi} = \frac{-p_{\perp}}{(1+\psi)}$ \cite{mora} is significantly reduced at the rear of the channel. The resulting electric field $E_z(\xi) = -\frac{d\psi_0}{dr_b}\frac{dr_b}{d\xi}$ therefore decreases from the rear to the front of the injected beam. The negative chirp $d_{\xi} E_z < 0$ is reproduced by the multi-sheath model (solid red) using $\psi_{min} = -0.93$ in Fig.~\ref{fig:injection_optimal}(b).

\begin{figure}[t]
\centering
\includegraphics[width=0.5\textwidth]{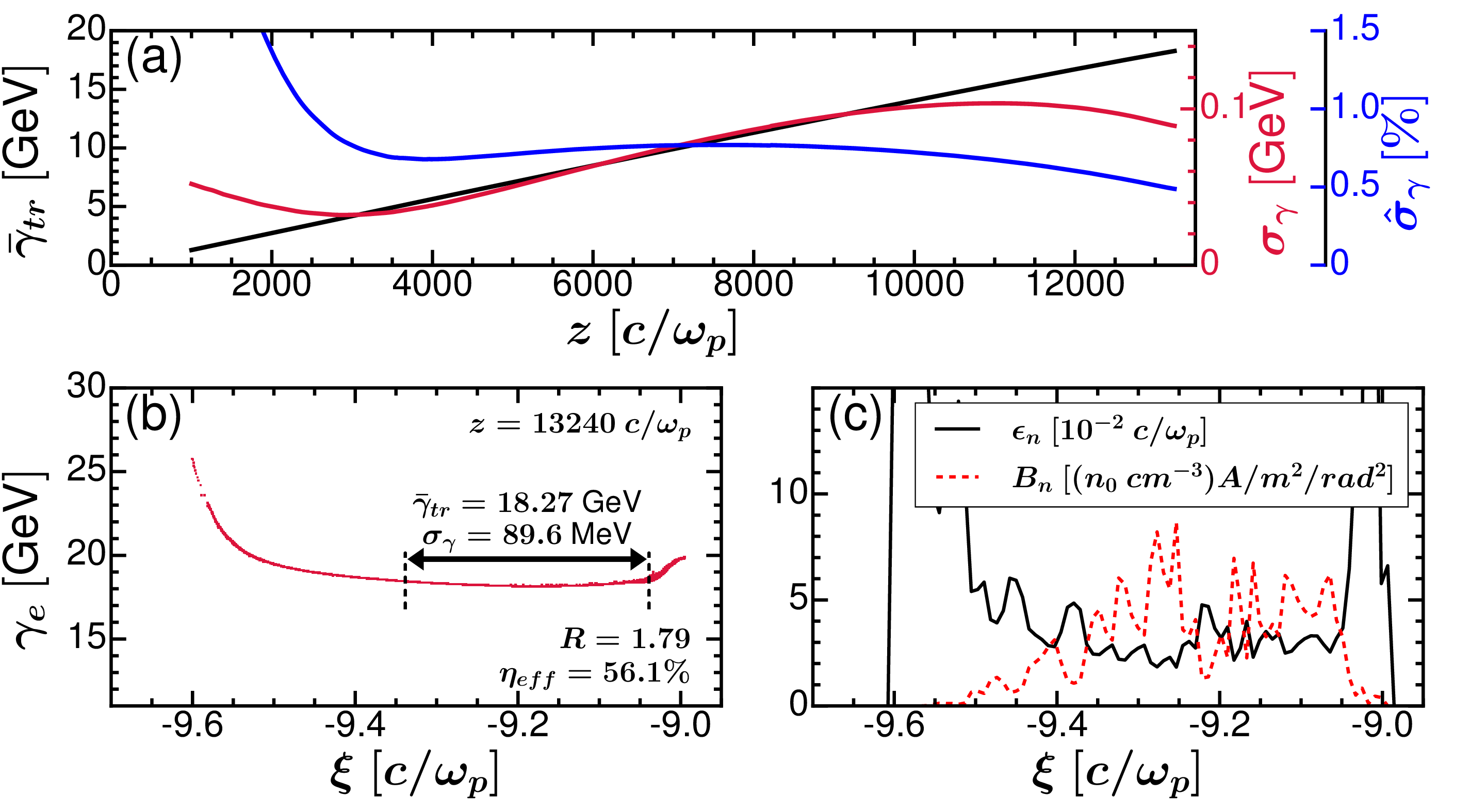}
\caption{\label{fig:duration_beams} (a) Average energy $\bar{\gamma}_{tr}$, absolute energy spread $\sigma_{\gamma}$, and normalized energy spread $\hat{\sigma}_{\gamma}$ of the trailing beam core as a function of distance $z$. (b) Energy-position phase space $(\gamma_e,\xi)$ and (c) injected beam slice parameters at $k_pz = 13240$. The beam core is denoted dashed black lines enclosing $50\%$ of the injected charge. }
\end{figure}

Fig.~\ref{fig:injection_optimal} shows that the electric field slope along the injected beam $d_{\xi}E_z$ decreases from positive to negative as the driver pump depletes. As a result, the average slope over the acceleration length $\langle d_{\xi} E_z \rangle \equiv \int_0^{L} dz^{\prime} (d_{\xi}E_z) / L $ decreases within the beam core as $L$ approaches the pump depletion length, $L_{pd}$. In Fig.~\ref{fig:duration_beams}(a), we plot the average energy $\bar{\gamma}_{tr}$, absolute energy spread $\sigma_{\gamma}$, and normalized energy spread $\hat{\sigma}_{\gamma}$ of the injected beam core as a function of propagation distance. Initially, $\sigma_{\gamma}$ and $\hat{\sigma}_{\gamma}$ decrease as the underloaded wakefield $(d_{\xi} E_z > 0)$ reduces the initial positive energy chirp of the beam following injection \cite{xu2017downrampinj}. A global minimum in $\sigma_{\gamma}$ therefore occurs near $3000 ~ c/\omega_p$ and $\sigma_{\gamma}$ subsequently increases as the beam acquires a negative energy chirp from the underloaded wakefield. It begins to plateau around $10000 ~ c/\omega_p$ and then decreases as the $d_{\xi}E_z$ shifts from underloaded to overloaded. After $12000 \ c/\omega_p$, the residual energy chirp rapidly reduces to sub-0.1 GeV and deviates significantly from the linear trendline of $\bar{\gamma}_{tr}$. This leads to $\hat{\sigma}_{\gamma} \sim \langle d_{\xi} E_z \rangle \sigma_z/\langle E_z \rangle $ approaching a global minimum at $z  = 13240 \ c/\omega_p$.

Figs.~\ref{fig:duration_beams}(b)-(c) show the final energy distribution and beam slice parameters at $z  = 13240 \ c/\omega_p$. Within the core of the beam (dashed black), we find an average energy of $18.27$ GeV, a projected energy spread of $89.6$ MeV (0.49\%), and slice energy spreads as low as $3$ MeV. Defining the transformer ratio, $R$, as the ratio of the average energy gain of the injected beam core to the maximum energy loss of the driver, we find $R \equiv \bar{\gamma}_{tr}/$max$(| \Delta \gamma_b |) \simeq 1.79$. A significant amount of charge $Q_{tr}/en_pk_p^{-3} \simeq 21.7$ is injected, corresponding to $0.52$ nC for $n_p = 10^{18} \centi \meter^{-3}$. A high transfer efficiency from the drive to trailing bunch $\eta_{eff}\equiv (\sum_{tr} q_{tr} \gamma_{tr})/(\sum_d q_d | \Delta \gamma_b |) \simeq 56.1 \%$ is achieved. Normalized slice emittances (solid black) as low as $0.11 \ \micro \meter$ ($0.02 \ c/\omega_p$) and normalized brightnesses (dashed red) as high as $9 \ [n_0 \ {\centi\meter}^{-3}]  \ \ampere/\meter^2/\rad^2$ are observed within the beam core. Additional simulations carried out using drivers with longer diffraction lengths up to $ 9600 \ c/\omega_p$ indicate that similar beams with core energies $\gtrsim 18$ GeV and core energy spreads $\hat{\sigma}_{\gamma} \lesssim 0.6 \%$ are produced. Simulations with $m=2$ out to $1000 ~c/\omega_p$ found the injection and initial acceleration was very similar.

\begin{figure}[b]
\centering
\includegraphics[width=0.5\textwidth]{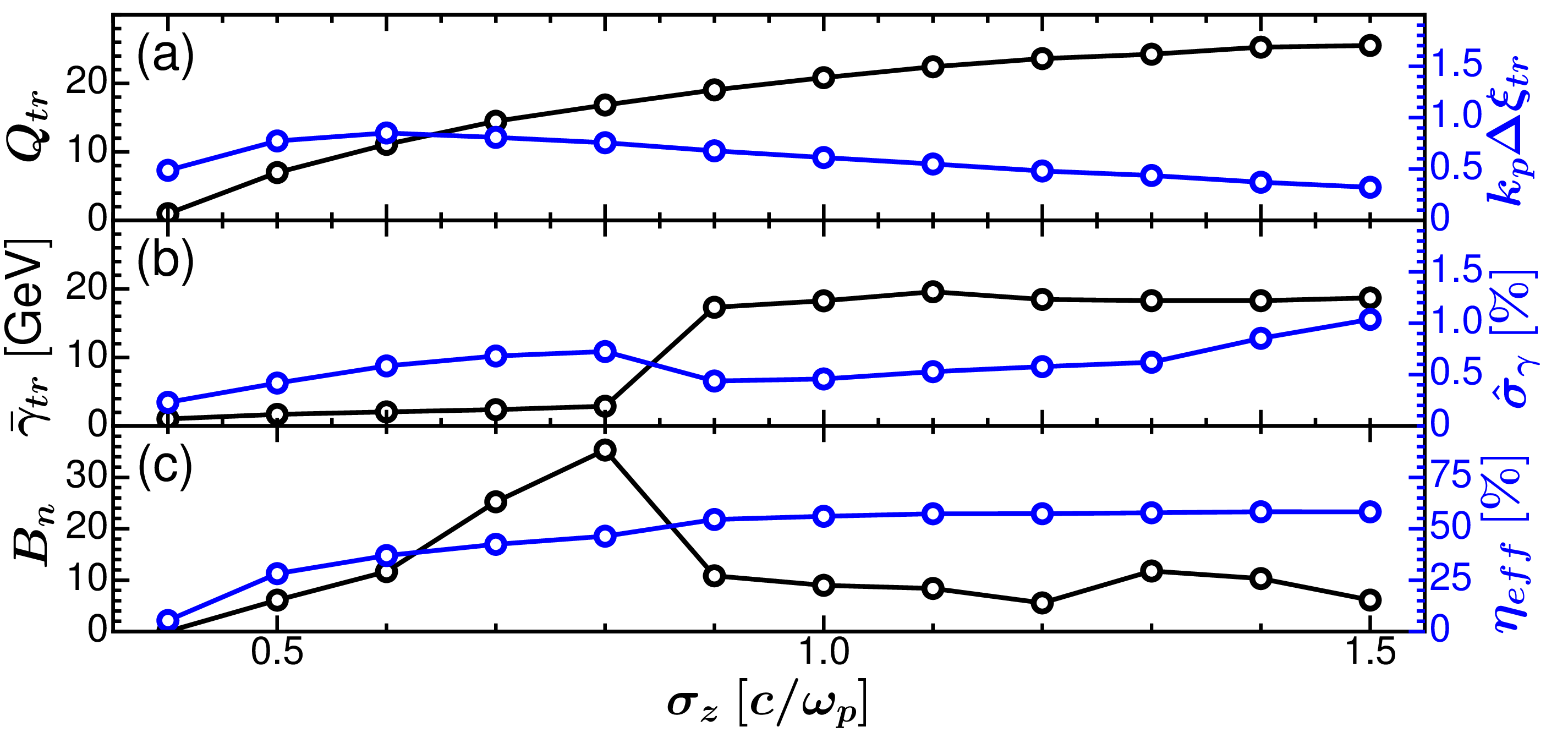}
\caption{\label{fig:duration_beams2} (a) Injected charge $Q_{tr}$ (units of $en_pk_p^{-3}$), beam length $\Delta\xi_{tr}$, (b) core energy $\bar{\gamma}_{tr}$, normalized core energy spread $\hat{\sigma}_{\gamma}$, (c) transfer efficiency $\eta_{eff}$ and peak normalized brightness $B_n$ (units of $[n_0 \ {\centi\meter}^{-3}]  \ \ampere/\meter^2/\rad^2$) for drivers $\{\gamma_b = 20000, \Lambda = 6, k_p\sigma_0 = 0.5\sqrt{\Lambda},k_{\beta}\beta^* = 16\}$ with different durations $k_p\sigma_z$. }
\end{figure}
 
Similar results are obtained for drivers with different durations, $k_p\sigma_z$, as seen in Fig.~\ref{fig:duration_beams2}. A clear transition to optimal beam loading occurs for $k_p\sigma_z \gtrsim 0.9$ due to higher slice currents $I\sim Q_{tr}/\Delta \xi_{tr}$ driving full rotation of $d_{\xi} E_z$ from negative to positive. For these parameters, the simulation results indicate that the injected beams can be accelerated to energies ranging from $17.3$--$19.6$ GeV with low core energy spreads $\hat{\sigma}_{\gamma} \lesssim 1\%$. In Fig.~\ref{fig:duration_beams2}(c), high transfer efficiencies (solid blue) in excess of $ 54\%$ and peak brightnesses (solid black) on the order of $\sim 10 \ [n_0 \ {\centi\meter}^{-3}]  \ \ampere/\meter^2/\rad^2$ are also observed for $k_p\sigma_z \gtrsim 0.9$. For plasma densities of $n_0 \sim 10^{18} \ \centi\meter^{-3}$, the results indicate that peak brightnesses on the order of $10^{19} \ampere/\meter^2/\radian^2$ can be achieved. These high-energy and bright beams may provide compact XFELs in the hundreds of keV photon range.

This work was supported by US NSF grant No. 2108970 and US DOE grant No. DE-SC0010064. The simulations were performed on the National Energy Research Scientific Computing Center (NERSC) and Hoffman2 at UCLA.

\bibliographystyle{apsrev4-1}
%reverse

%\raggedright
\bibliography{refs_thamine}
\end{document}